\documentclass[superscriptaddress, aps, prl, twocolumn, amsmath, amssymb, showpacs, nofootinbib]{revtex4-2}
\usepackage{graphicx}  
\usepackage{epstopdf}
\usepackage{dcolumn}   
\usepackage{bm}        
\usepackage{amssymb}   
\usepackage{amsmath}   
\usepackage{amsthm}
\usepackage{chemarrow}
\usepackage{color}
\usepackage[dvipsnames]{xcolor}
\usepackage{mathrsfs}
\usepackage{float}
\usepackage{physics}
\usepackage{slashed}
\usepackage{bbold}
\usepackage{bbm}
\usepackage{latexsym}
\usepackage{multirow}
\allowdisplaybreaks
\usepackage{hyperref}

\begin{document}

\title{Dark photon dark matter constraints at the Taiwan axion search experiment with haloscope}

\author{Yuan-Hann Chang}
\email{changy@gate.sinica.edu.tw}
\affiliation{Center for High Energy and High Field Physics, National Central University, Taoyuan City 320317, Taiwan}
\affiliation{Institute of Physics, Academia Sinica, Taipei City 115201, Taiwan}

\author{Cheng-Wei Chiang}
\email{chengwei@phys.ntu.edu.tw}
\affiliation{Department of Physics, National Taiwan University, Taipei 10617, Taiwan}
\affiliation{Physics Division, National Center for Theoretical Sciences, Taipei 10617, Taiwan}

\author{Hien Thi Doan}
\email{hiendoan86@gmail.com}
\affiliation{Department of Nuclear Physics, Faculty of Physics and Engineering Physics, University of Science, Ho Chi Minh city 700000, Vietnam}

\author{Nick Houston}
\email{nhouston@bjut.edu.cn}
\affiliation{Institute of Theoretical Physics, School of Physics and Optoelectronic Engineering, Beijing University of Technology, Beijing 100124, China}

\author{Jinmian Li}
\email{jmli@scu.edu.cn}
\affiliation{College of Physics, Sichuan University, Chengdu 610065, China}

\author{Tianjun Li}
\email{tli@itp.ac.cn}
\affiliation{School of Physics, Henan Normal University, Xinxiang 453007, P. R. China}
\affiliation{CAS Key Laboratory of Theoretical Physics, Institute of Theoretical Physics, Chinese Academy of Sciences, Beijing 100190, China}
\affiliation{School of Physical Sciences, University of Chinese Academy of Sciences, No. 19A Yuquan Road, Beijing 100049, China}

\author{Lina Wu}
\email{wulina@xatu.edu.cn}
\affiliation{School of Sciences, Xi'an Technological University, Xi'an 710021, P. R. China}

\author{Xin Zhang}
\email{zhangxin104774@lnnu.edu.cn}
\affiliation{Department of Physics, Liaoning Normal University, Dalian 116029, China}
\affiliation{Center for Theoretical and Experimental High Energy Physics, Liaoning Normal University, Dalian 116029, China}

\begin{abstract}
The dark photon is a well motivated candidate for the dark matter which comprises most of the mass of our visible Universe, leading to worldwide experimental and observational efforts towards its discovery.
A primary tool in this search is the cavity haloscope, which facilitates resonantly enhanced conversion to photons from both dark photons and axions.
In this context, limits from axion search experiments are often directly converted into dark photon constraints, without re-analyzing the original data. 
However, this rescaling may not fully capture all of the relevant physics due to various reasons.
By re-examining data taken by the Taiwan Axion Search Experiment with Haloscope (TASEH) experiment, we derive a world-leading constraint on the dark photon parameter space, excluding $|\epsilon|\gtrsim2\times10^{-14}$ in the $19.46 - 19.84\,\mu$eV mass range, which exceeds the na{\"i}ve `rescaling limit' by roughly a factor of two.
We emphasize that accounting for the scanning timing information is crucial for deriving limits for the polarized dark photon case.
In the data, we also analyze a tentative signal excess with a local significance of 4.7$\sigma$ ($m_X \simeq 19.5\,\mu$eV) that persists in the absence of a magnetic field. 
While this excess mimics the behavior of a dark photon signal, it has been excluded by recent results from the HAYSTAC and ORGAN-Q experiments. 
This case study, nevertheless, highlights the risk of discarding valid dark photon signals when relying on axion-specific magnetic field vetoes.
\end{abstract}

\maketitle

\noindent \textbf{Introduction.} 
The dark photon is a hypothetical massive vector boson, commonly occurring in some minimal gauge extensions of the Standard Model (SM). 
It provides a naturally light dark matter (DM) candidate which kinetically mixes with the SM photon, creating interesting phenomenology and facilitating a wide variety of detection schemes \cite{Fabbrichesi:2020wbt}.
As a result, intense experimental efforts are ongoing worldwide in dark photon searches.

A primary tool in such searches is the cavity haloscope \cite{Brubaker:2016ktl, ADMX:2018gho, Nguyen:2019xuh, HAYSTAC:2020kwv, CAPP:2020utb, AlvarezMelcon:2020vee,CAST:2020rlf,Grenet:2021vbb,Cervantes:2022epl, Cervantes:2022gtv, McAllister:2022ibe, Cervantes:2022yzp, ALPHA:2022rxj,Schneemann:2023bqc, SHANHE:2023kxz, APEX:2024jxw, He:2024fzj,QUAX:2024fut,MADMAX:2024sxs}, originally developed as an experimental technique to search for the axion \cite{Sikivie:1983ip}, another light DM candidate which can convert to photons. 
Cavity haloscopes enhance the sensitivity to certain processes in a narrow frequency range, at the expense of broadband sensitivity, thereby allowing weakly-interacting DM to potentially be detected in the laboratory.
Axion-photon conversion proceeds via the Primakoff effect, relying on a background magnetic field.
Conversion of dark photons to photons meanwhile is also resonantly enhanced by cavity haloscopes, with the benefit of not requiring a magnetic field.

Many axion constraints have already been reinterpreted to render novel dark photon constraints \cite{Caputo:2021eaa}.
However, care has to be taken as na{\"i}ve rescalings may not correctly capture all the relevant physics.
For example, whilst the axion is a pseudoscalar, the dark photon is a spin-one gauge field and hence has a polarization, the orientation of which relative to a detection apparatus can have significant effects on the signal strength.
Since this orientation is in general unknown, with the precise details of the dark photon polarization dependent on cosmological history, any derived limit on new physics must correctly account for this additional uncertainty.

Furthermore, many axion detection schemes rely on a magnetic field veto to exclude spurious signals, whose power does not vary in proportion to the magnetic field strength.
Whilst effective in the axion context, dark photon conversion does not require a magnetic field, and hence such a veto would have no effect on a dark photon signal.
This may potentially lead to valid candidate signals being discarded.

With these points in mind, in this work we re-examine data taken by the Taiwan Axion Search Experiment with Haloscope (TASEH), which provides the strongest limits on axion DM in the 19.46 to 19.84 $\mu$eV mass range, excluding axion-photon couplings $|g_{a\gamma\gamma}|\gtrsim8.2\times10^{-14}$ GeV${}^{-1}$ at 95\% confidence level (C.L.)~\cite{TASEH:2022vvu,TASEH:2022noe}. 

This axion constraint has previously been rescaled to provide a corresponding dark photon DM limit ~\cite{AxionLimits}. 
However, as we will demonstrate, this `rescaled' bound is overly conservative in the case of polarized dark photon DM, due to the lack of timing information and the assumption of an instantaneous measurement.
By incorporating timing data recorded by TASEH, we derive a more precise and stringent bound for polarized dark photon DM. 
Our analysis follows the methods outlined in Ref.~\cite{Caputo:2021eaa}, providing a practical example of converting axion to dark photon constraints.

During the TASEH measurement, an excess with a signal-to-noise ratio (SNR) exceeding $3\sigma$ was also observed in the 4.71017 to 4.71019~GHz range.
This excess persisted after 14 repeated scans and could not be attributed to any known sources. 
Since the signal strength was unaffected by changes in the external magnetic field, it was not believed to originate from axions, and the corresponding frequency region was excluded from the original analysis. 
However, such an excess behaves phenomenologically like a tentative dark photon signal.

We perform a detailed analysis of this excess to demonstrate how dark photon signals could be easily overlooked in standard axion analyses.
While the data collected by TASEH is consistent with a dark photon DM hypothesis with kinetic mixing of $6.48 \times 10^{-15}$ at a mass of 4.710178~GHz (local significance 4.76 $\sigma$), we note that this parameter space has been excluded recently by the HAYSTAC~\cite{Bai:2025fpr} and ORGAN-Q~\cite{Quiskamp:2025wme} experiments.
This strongly suggests the excess is an instrumental artifact.
Nonetheless, the analysis validates our methodology and reinforces the importance of dedicated dark photon searches which do not rely on magnetic field vetos.

\noindent \textbf{Dark photon cavity searches}. 
The low-energy effective Lagrangian for a dark photon kinetically mixing with the SM photon can be given by
\begin{align}
\mathcal{L} 
=& 
-\frac{1}{4} F_{\mu \nu} F^{\mu \nu} -\frac{1}{4} X_{\mu \nu} X^{\mu \nu} +\frac{\sin \alpha}{2} F^{\mu \nu} X_{\mu \nu}
\nonumber \\
&+ e J^\mu_{\text{EM}} A_\mu -\frac{m^2_X \cos^2 \alpha}{2} X^\mu X_\mu\,,
\end{align} 
where $F_{\mu \nu}$ and $X_{\mu \nu}$ are respectively the field strength tensors of the SM photon and the dark photon, $J^{\mu}_{\text{EM}}$ is the electromagnetic current, $\sin \alpha$ is the kinetic mixing parameter, and $m_X$ is the mass of the dark photon $X$. 
The signal power due to the dark photon DM in a cavity haloscope is given by
{
\begin{equation}
P_{\rm dp} = \epsilon^2 m_X \rho_{\text{DM}} V C^X_{010} Q_L  \frac{\beta}{1+\beta} L(f,f_c,Q_L), \label{eq:powerdp}
\end{equation}
where $\epsilon=\tan \alpha$ is the coupling strength,  $\rho_{\text{DM}} \simeq 0.45$ GeV/cm$^{3}$ is the DM density, $V$ is the cavity volume, $Q_L$ is the loaded quality factor, and $\beta$ is the cavity coupling coefficient. 
The Lorentzian term~\cite{Caputo:2021eaa}, 
\begin{equation}
\label{eq::Lorentzian}
L(f,f_c,Q_L)=\frac{1}{1+4 Q_L^2 (f/f_c -1)^2}
\end{equation}
is a detuning factor with $f$ being the SM photon frequency and $f_c$ being the cavity resonance frequency. 
The form factor $C^X_{010}$ is the normalized overlap between the dark photon and the induced electric field mode $\textbf{E}_{010}(\vec{x})$
\begin{equation}
C^X_{010} 
= 
\frac{(\int dV \textbf{E}_{010}(\vec{x}) \cdot \hat{X}(\vec{x}))^2}{V \int dV |\textbf{E}_{010}(\vec{x})|^2 |\hat{X}(\vec{x})|^2 }\,,  
\label{eq:geo}
\end{equation}
with $\hat{X}$ being the dark photon polarization orientation.
}


Following the analysis laid out in Ref.~\cite{Caputo:2021eaa}, we work in geocentric equatorial coordinates, where
\begin{equation}
\hat{X} = (\sin \theta_X \cos \phi_X, \sin \theta_X \sin \phi_X, \cos \theta_X) \,,
\end{equation}
while the sensitive direction of the cavity (since the cavity is ``zenith-pointing") is 
\begin{equation}
\hat{Z}(t) = (\cos \lambda_{\text{lab}} \cos \omega t, \cos \lambda_{\text{lab}} \sin \omega t, \sin \lambda_{\text{lab}}) \,,
\end{equation}
with $\lambda_{\text{lab}}$ denoting the latitude and the angular frequency of the Earth's rotation $\omega =2 \pi / (1~\text{sidereal}~\text{day})$. 
Then the angle of interest is $\cos \theta(t) = \hat{X} \cdot \hat{Z}(t)$.

The dark photon DM signal power accumulated over a measurement time $T$ is proportional to the time-averaged 
\begin{equation}
\langle \cos^2 \theta(t) \rangle_T = \frac{1}{T} \int^T_0 \cos^2 \theta(t) dt
\label{eq:cost1} 
\end{equation}
for multiple measurements with start times $T^{\text{start}}_i$ and durations $T_i$ ($i=1, \cdots, N$).  In this case,
\begin{eqnarray}
\langle \cos^2 \theta(t) \rangle_T 
=& 
\frac{1}{\sum P^0_i} \bigg( \frac{P^0_1}{T_1} \int^{T^{\text{start}}_1+T_1}_{T^{\text{start}}_1} \cos^2 \theta(t) dt
\nonumber \\
&+  \frac{P^0_2}{T_2} \int^{T^{\text{start}}_2+T_2}_{T^{\text{start}}_2} \cos^2 \theta(t) dt + \cdots \bigg) \,,\label{eq:cost2} 
\end{eqnarray}
where $P^0_i$ is the dark photon power given in Eq.~\eqref{eq:powerdp}. 

For an unpolarized dark photon DM, $(\theta_X, \phi_X)$ are sampled isotropically across the sky. 
Given that 
\begin{equation}
\frac{1}{4 \pi} \int \langle \cos^2 \theta(t) \rangle_T d \cos \theta_X d\phi_X = \frac{1}{3} \,,
\end{equation}
regardless of the measurement time, one can conclude that the conversion factor $\langle \cos^2 \theta(t) \rangle_T$ is always 1/3 for the unpolarized case. 

For a dark photon with an unknown fixed polarization, assuming an isotropic prior, one can calculate the 95\% C.L. limit on the signal power via 
\begin{align}
\int  dP_X 
&\int dN  f(P_X) f(N) =\int_0^1 d \langle \cos^2 \theta(t) \rangle_T  \frac{f(\langle \cos^2 \theta(t) \rangle_T)}{2}
\nonumber \\
&\times \left[1+ \text{erf}\left(\frac{- P^0_X \langle \cos^2 \theta(t) \rangle_T }{\sqrt{2} \sigma_N}\right) \right]=0.05\,, 
\label{eq:conv1}
\end{align}
where $\sigma_N$ is the standard deviation of the noise, and
$f(\langle \cos^2 \theta(t) \rangle_T)$ is the probability distribution of the time-averaged $\langle \cos^2 \theta(t) \rangle_T$, which is obtained by sampling $(\theta_X, \phi_X)$ in Eqs.~\eqref{eq:cost1}-\eqref{eq:cost2} isotropically. 
As a result, the conversion factor for a 95\% C.L. limit on signal power can be expressed as
\begin{equation}
 \langle \cos^2 \theta(t) \rangle_{T}^{\text{excl.}}= \frac{\Phi^{-1}[0.95] \sigma_N}{P^0_X} \,, \label{eq:convfactor}
\end{equation}
with $\Phi^{-1}$ denoting the inverse cumulative distribution function of the Gaussian noise.

\noindent \textbf{The TASEH experiment.} TASEH is located at the Department of Physics, National Central University, Taiwan~\cite{TASEH:2022vvu,TASEH:2022noe}.
In the present analysis, we consider the CD102 dataset, which was collected by TASEH from October 13, 2021 to November 15, 2021.

The TASEH cavity geometry is a cylinder split into two halves along the axial direction with an effective volume of 0.234~L, made of oxygen-free high-conductivity (OFHC) copper and housed within a cryogen-free BlueFors dilution refrigerator (DR).
An 8-Tesla superconducting solenoid with a bore diameter of 76~mm and a length of 240~mm is also integrated with the DR.

The resonant frequency of the TM$_{010}$ mode at cryogenic temperatures is tuned over the range of 4.70750 to 4.79815~GHz via rotation of an off-axis OFHC copper tuning rod.
Each tuning step is approximately 95 to 115~kHz. 

The values of C$_{010}$, {for the axion case}, are derived from the magnetic field map provided by BlueFors and cavity electric field distributions simulated with Ansys HFSS (high-frequency structure simulator).

During data taking, the temperature of the cavity was $T_c \simeq 155$~mK.
The overall system noise is dominated by the effective temperature of the noise added by the receiver chain, $T_a \simeq 2$~K. 
The DR temperature typically rose by a few mK due to the operations of the rotational motor during tuning. However, by the time data acquisition began, the DR temperature had dropped back to the base temperature $T_{\text{mx}} = 27$~mK. 
Therefore, this temperature rise did not affect data taking.

Table~\ref{tab:taseh} provides benchmark values for key experimental parameters, including the intrinsic, unloaded quality factor $Q_0$ at cryogenic temperatures, the number of tuning steps $N_{\text{step}}$, and the resolution bandwidth used in the data analysis $\Delta f_c$. 
More details regarding measurements of some of these parameters may be found in Ref.~\cite{TASEH:2022hfm}.

\begin{table}[htbp]
	\centering 
	\scalebox{0.85}{
	\begin{tabular}{|c|c|c|c|c|c|c|} \hline
     $f_{\text{low}}$(GHz) & $f_{\text{high}}$(GHz) & $N_{\text{step}}$ & $\Delta f_{\text{step}}$(kHz) & $B_0$(T) & $V$(L) & $C_{010}$ \\ \hline
     4.70750 & 4.79815 & 837 & 95-115 & 8 & 0.234 & 0.6 \\ \hline \hline
     $Q_0(\times10^4)$ & $\beta$ & $T_{\text{mix}}$ (mK)& $T_c$ (mK) & $T_a$ (K) & $\Delta f_c$ (kHz) & $\lambda_{\text{lab}}$ \\ \hline
     5.8-6.5 & 1.9-2.3 & 27-28& 155 & 1.9-2.2 & 1 & $25^{\circ}$ \\ \hline
	\end{tabular}}
	\caption{\label{tab:taseh} The key detector parameters for TASEH. }
\end{table}

Each tuning step is referred to as a ``scan.'' 
The frequency span for each scan is 1.6~MHz, and the tuning frequency steps are $\sim 100$~kHz. 
Therefore, each frequency within the range of the experiment was scanned $15-16$ times during the main data-taking run. 
The integration time of each scan was about $32-42$ minutes, designed to achieve an SNR of approximately 3.36. 

A Savitzky-Golay (SG) filter is performed on the averaged spectrum of each frequency scan, using a window width of 201 frequency bins. 
In the absence of a signal, the SG filter output can be considered as the averaged noise power. 
The raw averaged power spectrum is divided by the output of the SG filter, then a unity is subtracted to get the dimensionless normalized spectrum, called the Relative Deviation of Power (RDP), denoted by $\delta^{\text{norm}} = \text{(Raw Power)}/\text{(SG filter)} -1$. 
The normalized spectrum from each scan is further rescaled such that a KSVZ axion signal will have a rescaled RDP $\delta^{\text{res}}_{ij}$ that is approximately equal to unity, if the signal power is contained within only one frequency bin:
\begin{eqnarray}
\delta^{\text{res}}_{ij}&=R_{ij} \delta_{ij}, ~ \sigma^{\text{res}}_{ij}=R_{ij} \sigma_i ~,~  \text{for $i$-th scan, $j$-th bin}\,,\nonumber\\
&{\rm where}\,\,R_{ij}= \frac{k_B T_{\text{sys}} \Delta f_{\text{bin}}}{P^{\text{KSVZ}}_{ij} L(f_{ij},f_{ci},Q_L)}\,,
\end{eqnarray}
with $f_{ci}$ denoting the cavity frequency of $i$-th scan. 
The system noise temperature $T_{\text{sys}} \sim T_a$ to a good approximation, where $T_a$ can be estimated from data fitting. 

To combine spectra from different resonant-frequency scans, thereby obtaining a single overall spectrum, one needs to apply a weighting algorithm. 
So, for a given frequency $f_{n}$, we define the combined spectrum as 
\begin{eqnarray}
\delta^{\text{com}}_n =\frac{\sum ( \delta^{\text{res}}_{ij} \cdot \omega^{ij}_n)}{ \sum  \omega^{ij}_n}, \,\,\sigma^{\text{com}}_n = \frac{\sqrt{\sum (\sigma^{\text{res}}_{ij} \cdot \omega^{ij}_n )^2} }{\sum \omega^{ij}_n} \,, \label{eq:combrdp}
\end{eqnarray}
where $\omega^{ij}_n = 1/{ (\sigma^{\text{res}}_{ij} )^2}$, and summations are over the $i$-th scan, $j$-th bin matching the target frequency $f_{n}$.

Bins in the combined spectrum are then combined to create the final merged spectrum, from which axion and dark photon limits are derived.
The precise details of this merging are not essential to our analysis here and are provided in the Supplemental Material.
A more detailed description of the analysis procedure can be found in Ref. ~\cite{TASEH:2022noe}.

\noindent \textbf{Dark photon constraints}.  
We first note that the axion power in a cavity haloscope is
\begin{equation}
P_{\rm a} = \left(\frac{g^2_{a\gamma} B_0^2}{m_a}\right)  \rho_{\text{DM}} V C_{010} Q_L  \frac{\beta}{1+\beta} L(f,f_c,Q_L)~, \label{axionpower}
\end{equation}
where $g_{a\gamma}$ is the axion-photon-photon coupling and $B_0$ is the nominal strength of the external magnetic field.
By comparison with the dark photon power in Eq.~\eqref{eq:powerdp}, we can obtain a naive rescaling between the axion coupling and the dark photon kinetic mixing, $\epsilon = g_{a\gamma} \frac{B_0}{m_X |\cos \theta|}$, where $\theta$ is the angle between the dark photon polarization $\hat{X}$ and the external magnetic field.

However, to more correctly incorporate the dark photon polarization, we need to use the time-averaged $ \langle \cos^2 \theta(t) \rangle_{T}$.  
By requiring the dark photon limit to have the same significance as the axion limit, we obtain the following relation for the SNR 
\begin{eqnarray}
1.64 (\sim 95\%\text{C.L.}) 
&= 
\left( \frac{p^0_X \epsilon^2 m_X \langle \cos^2 \theta(t) \rangle^{\text{excl.}}_T}{\sigma_N} \right)_{\text{DP}} 
\nonumber\\
 &= \left( \frac{p^0_X g^2_{a\gamma} B_0^2 }{m_a \sigma_N} \right)_{\text{Axion}} ~,~
\end{eqnarray}
where $p^0_X$ represents the shared term in the signal powers for both dark photon and axion, as given in Eqs.~(\ref{eq:powerdp}) and~(\ref{axionpower}).
As a result, the corresponding axion coupling limit can be translated to a kinetic mixing limit as follows:
\begin{align}
\epsilon &= g_{a\gamma} \frac{B_0}{m_X \sqrt{\langle \cos^2 \theta(t) \rangle^{\text{excl.}}_T}} ~,~ 
\end{align}
 where the $\langle \cos^2 \theta(t) \rangle^{\text{excl.}}_T$ is the conversion factor describing how much the dark photon power threshold needs to be enhanced over the axion case for a given exclusion.

\begin{figure}[htb]
\begin{center}
\includegraphics[width=0.48\textwidth]{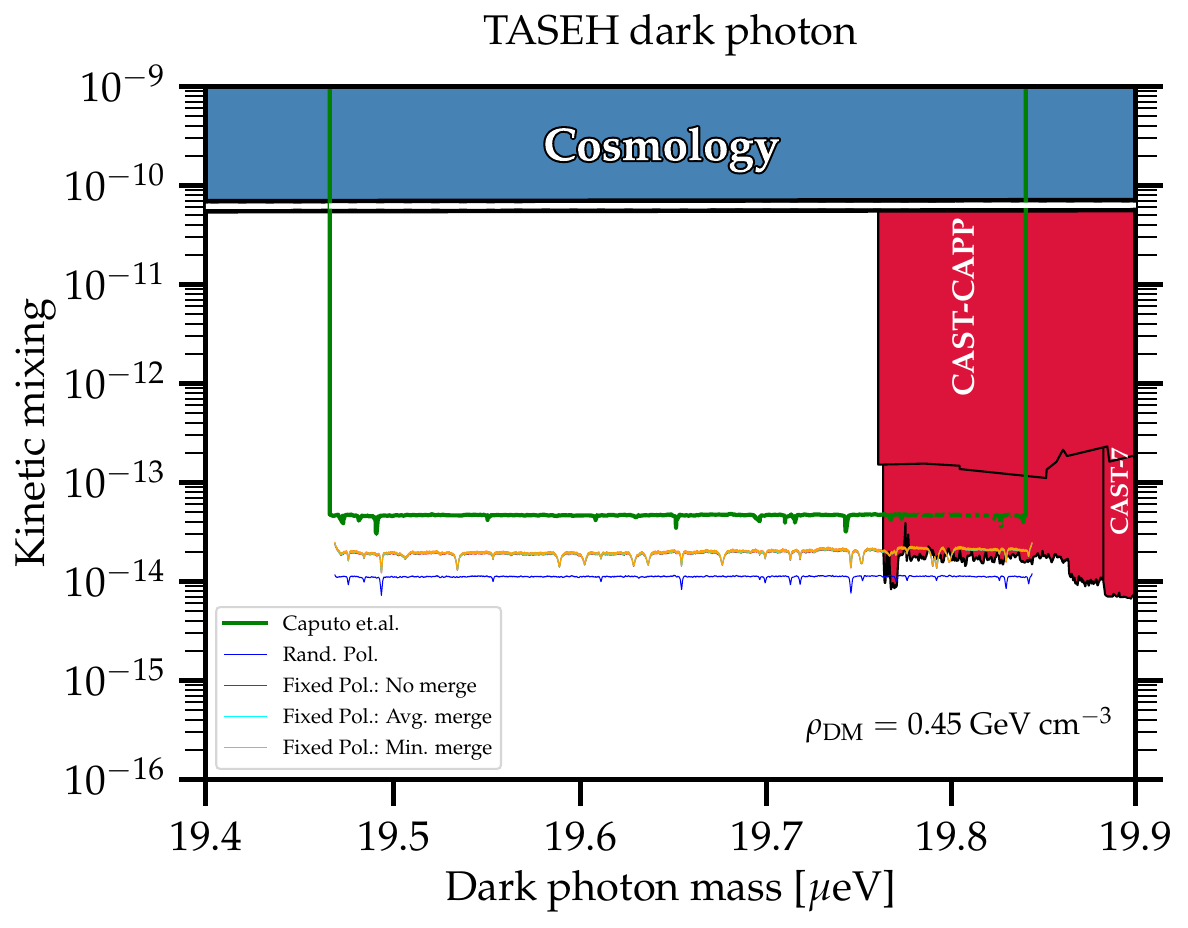}
 \end{center}
\caption{
Constraints on the dark photon parameter space from TASEH. 
The green curve corresponds to the AxionLimits constraint \cite{AxionLimits}, which follows the analysis strategy presented in \cite{Caputo:2021eaa}. 
The blue bound corresponds to the random polarization case. 
{The red, cyan, and orange curves respectively denote the bounds for polarized dark photon DM without bin merging and with merged bins using the average or minimal conversion factors; their overlap indicates the minimal effect of the merging procedure.}
\label{fig::dpbound}}
\end{figure}

Figure~\ref{fig::dpbound} presents our resulting bounds on the dark photon parameter space. 
Note that in Ref.~\cite{Caputo:2021eaa} a very short measurement time was typically assumed for cavity haloscopes, which resulted in a value of $\sim 0.02$ for the conversion factor.
Since the true $\langle \cos^2 \theta(t) \rangle^{\text{excl.}}_T \sim 0.1$ at TASEH, these assumptions weaken the bound on $|\epsilon|$.

Our result improves upon other limits by around four orders of magnitude in the mass range of [19.5,19.8]$~\mu$eV. 
The limit for the randomly polarised case is a factor of $\sim 2$ times stronger than the fixed polarization case, since the conversion factor in the former is 1/3 irrespective of the measurement time. 
Moreover, compared to the AxionLimits constraint \cite{AxionLimits} which uses direct rescaling, our constraint is stronger by a factor around 2.

\noindent \textbf{Analysis of non-magnetic excess}. 
As discussed previously, axion cavity haloscopes often rely on a magnetic field veto to exclude spurious signals, which may result in genuine dark photon signals being erroneously excluded.

In the CD102 dataset, there were two frequency ranges with an SNR greater than 3.355 after rescans. 
The signal around 4.7473~GHz was judged to have originated from an instrument control computer in the laboratory, and so the limit on this frequency is not included in Figure~\ref{fig::dpbound}.  
Interestingly, a signal in the range of $4.71017$ – $4.71019$~GHz existed only inside the DR, and persisted even after turning off the magnetic field. 
Although this excess is not an axion signal, it could be misinterpreted as a genuine dark photon signal if one does not cross-check with other experiments.

Including rescans, the frequency range 4.71017 – 4.71019~GHz was scanned 28 times, with measurement times spread out over the sidereal day. 
In principle, we could extract the dark photon polarization by correlating the signal strength with the detector orientation relative to the DM halo.
However, the SNR is low for each individual scan, even when the cavity is close to resonance with the signal frequency. 
As such, there is only a weak sensitivity to the dark photon polarization. 
Since the combined significance is nonetheless sizeable, we perform the analysis in the random polarization scenario.


The fitting process involves the minimization of $\Delta \chi^2$, which quantifies the discrepancy between observed data and model predictions. 
The $\Delta \chi^2$ is defined as 
\begin{equation}
\Delta \chi^2=\sum_i\left(\frac{\delta^{\text{com}}_i - \mu^2 \int_i F(f,f_X) df}{\sigma^{\text{com}}_i}\right)^2~,~
\end{equation}
where the fitting parameters are the signal strength $\mu$ and the dark photon frequency $f_X$, and $F$ is the DM lineshape (see Supplemental Material for details). 
The integration $\int_i F(f,f_X) df$ is performed over the range of the $i$-th frequency bin. 
$\Delta \chi^2$ is calculated by summing over 120 frequency bins centered around 4.710180~GHz. 
Before including rescan data, the best-fit signal strength is $\mu \equiv \left( {\epsilon}/{\epsilon_0} \right)^2=101.68$ (corresponding to a kinetic mixing $|\epsilon|=1.013 \times 10^{-14}$) at $f_{X}=4.710178$~GHz with $\Delta \chi^2 =12.82$.  
After rescans, the best-fit signal strength is slightly reduced to $\left( {\epsilon}/{\epsilon_0} \right)^2=41.59$ (corresponding to a kinetic mixing $|\epsilon|=6.48 \times 10^{-15}$) at the same frequency with $\Delta \chi^2 =21.40$.
However, the standard deviation is also reduced by a factor of two after rescanning, so the local signal significance is increased from 3.755 to 4.762.
To estimate the global significance, we follow the approach taken in Ref.~\cite{Foster:2017hbq} to estimate the trials factor. 
Due to the very large number of trial frequencies, we find a global significance of 1.729$\sigma$ after rescanning.

\begin{figure}[htbp]
\begin{center}
\includegraphics[width=0.45\textwidth]{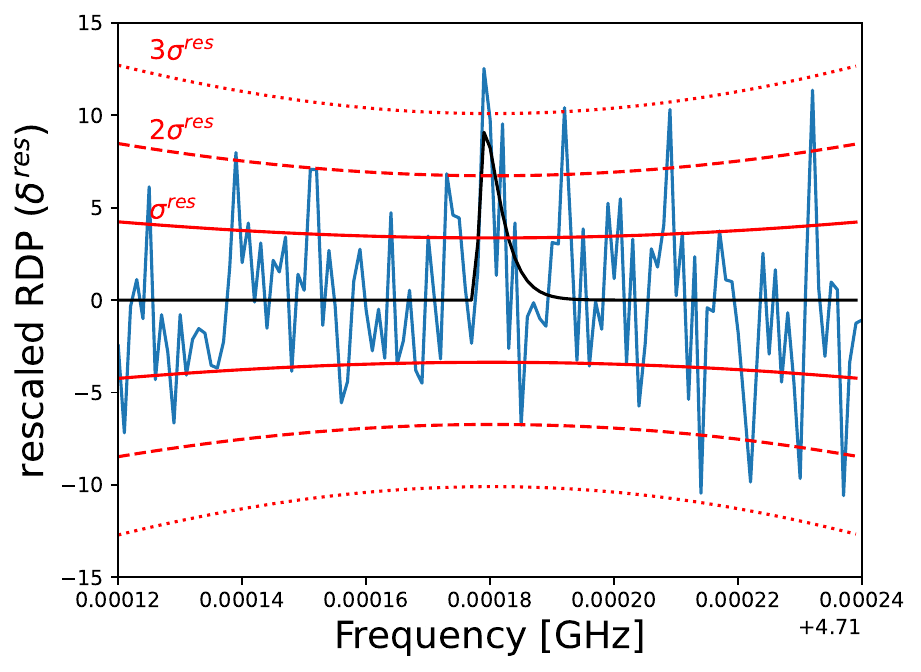}
\end{center}
\caption{
Rescaled RDP for the combined spectrum after rescans in the frequency range $[4.710120,4.710240]$ GHz. 
Red lines correspond to the RDP standard deviation, while the dark curve corresponds to the best-fit dark photon lineshape. 
\label{fig::rdp}}
\end{figure}
We plot the signal lineshape in terms of rescaled RDP after rescans in Figure~\ref{fig::rdp}, together with the RDP standard deviation.  
However, we note that the HAYSTAC and ORGAN-Q experiments have recently probed this frequency range with superior sensitivity and found no evidence for a signal. 
This external information leads us to conclude that the excess in TASEH data is likely an instrumental artifact rather than a genuine dark photon signal. 
This example clearly illustrates the challenges in dark photon searches where the magnetic field veto cannot be used to distinguish signal from noise, and emphasizes the need for multiple independent experiments to cross-check potential candidates.



\noindent \textbf{Conclusions and discussions}. 
The dark photon is a well-motivated DM candidate, with ongoing efforts worldwide into its discovery.
A primary technique in this search is the cavity haloscope, which permits resonantly enhanced conversion to photons from both dark photons and axions, allowing laboratory searches for weakly interacting DM.
Given certain phenomenological similarities between dark photon and axion DM, limits from axion searches in this context are often directly converted into dark photon constraints, without re-analyzing the underlying data. 
However, due to various reasons argued above, this na{\"i}ve rescaling may not fully capture all of the relevant physics.

By re-examining data taken by the Taiwan Axion Search Experiment with Haloscope (TASEH) experiment, we have derived a world-leading constraint on the dark photon parameter space, excluding $|\epsilon|\gtrsim2\times10^{-14}$ in the $19.46 - 19.84\, \mu$eV mass range, exceeding the `rescaling limit' by roughly a factor of two. This significant improvement underscores the importance of correctly modeling the dark photon polarization and incorporating detailed timing information from the experiment.

In addition, we identified a tentative signal excess with a local significance of 4.7$\sigma$ ($f_X\simeq 4.71$~GHz) which persisted in the absence of the magnetic field.
While such a signal would be vetoed in an axion search, it mimics the signature of a dark photon. 
However, given that this candidate region has been excluded by recent results from HAYSTAC and ORGAN-Q, we attribute this excess to spurious noise.
This analysis serves as a valuable case study, demonstrating that direct rescaling of axion constraints to dark photon may suffer significant inaccuracies, and that the dedicated reanalyses are essential for robust dark photon limits.
Therefore, other experimental Collaborations may also wish to perform their own dedicated reanalyses.

\noindent \textbf{Acknowledgments.}
CWC is supported by the National Science and Technology Council under Grant No.~NSTC-111-2112-M-002-018-MY3 and in part by a grant from the Simons Foundation (1161654, Troyer) for the Aspen Center for Physics.
TL is supported in part by the National Key Research and Development Program of China Grant No. 2020YFC2201504, by the Projects No. 11875062, No. 11947302, No. 12047503, and No. 12275333 supported by the National Natural Science Foundation of China, by the Key Research Program of the Chinese Academy of Sciences, Grant No. XDPB15, by the Scientific Instrument Developing Project of the Chinese Academy of Sciences, Grant No. YJKYYQ20190049, and by the International Partnership Program of Chinese Academy of Sciences for Grand Challenges, Grant No. 112311KYSB20210012.
NH is supported by the National Natural Science Foundation of China, Grant No. 12150410317, and the Beijing Natural Science Foundation, Grant No. IS23025. 
JL is supported by the National Natural Science Foundation of China under Grant No. 11905149, by the Natural Science Foundation of Sichuan Province under Grant No. 2023NSFSC1329. 
LW is supported in part by the Natural Science Basic Research Program of Shaanxi, Grant No. 2024JC-YBMS-039 and No. 2024JC-YBMS-521.

\newpage
\onecolumngrid

\appendix


\section{Appendix A: Polarised dark photon production in the early Universe}
Several studies~\cite{Graham:2015rva,Ema:2019yrd,Kolb:2020fwh,Ahmed:2020fhc,Agrawal:2018vin,Co:2018lka,Bastero-Gil:2018uel,Alonso-Alvarez:2019ixv,Long:2019lwl,Nakai:2020cfw,Co:2021rhi,Cyncynates:2023zwj,Cyncynates:2024yxm} have proposed production mechanisms for light dark photon DM in the early Universe. 
Whilst these mechanisms are all capable of producing the required DM relic abundance, they do differ in the characteristics of the dark photon DM they produce.

For example, the polarization can be fixed within the entire cosmological horizon, the so-called fixed polarization scenario.
Alternatively, it is also possible that the dark photon field lacks any net polarization.
This is the so-called random polarization scenario, where the polarization varies randomly over timescales longer than the coherence time set by the corresponding dark photon de Broglie wavelength.


Dark photon DM can be produced in the early Universe via quantum fluctuations during inflation~\cite{Graham:2015rva,Ema:2019yrd,Kolb:2020fwh,Ahmed:2020fhc}. 
The production of the longitudinal modes requires $0<m_X < H_I$, where $H_I$ is the typical inflationary scale.
The resulting relic density of the longitudinal dark photon mode can then be given by $\Omega_X/\Omega_{\text{CDM}} = \sqrt{\frac{m_X}{6 \times 10^{-6} ~\text{eV}}} (\frac{H_I}{10^{14}~\text{GeV}})^2$~\cite{Graham:2015rva}. 
The inflationary production of transverse modes requires an extra coupling of the inflaton field to the dark photon, along with a relatively small inflationary scale~\cite{Beaufort:2023qpd}. 
Note that these two possibilities are mutually exclusive: for a given choice of dark photon mass and cosmological parameters, inflationary production will only result in either longitudinal or transverse dark photon polarization in the DM population. 

Similar to axion DM~\cite{Preskill:1982cy,ABBOTT1983133,Dine:1982ah}, dark photon DM can also be produced via the misalignment mechanism. 
However, in contrast to the axion, a minimal coupling to gravity causes the dark photon to behave like radiation in the early Universe (\textit{i.e}, $\rho \propto R^{-4}$ for $t \ll m_X^{-1}$), leading to an under-abundant relic density. 
A non-minimal coupling to the Ricci scalar $\frac{\kappa}{6} R X_\mu X^\mu$ with $\kappa\sim \mathcal{O}(1)$ can be introduced to alleviate this tension~\cite{Arias:2012az,Alonso-Alvarez:2019ixv}. 
The misalignment production mechanism naturally leads to the fixed polarization scenario.

Another dark photon production scenario is based on tachyonic instabilities~\cite{Agrawal:2018vin,Co:2018lka,Bastero-Gil:2018uel}. 
Here the energy density is initially stored in the zero mode of the axion field, which is later transferred to the transverse and longitudinal components of the dark photon. 
The relic density of the dark photon in this case can be written as 
\begin{equation}
    \Omega_{A^\prime} h^2  \simeq 0.2\, \theta^2 \left(\frac{40}{\beta'} \right) \left(\frac{m_{X}}{10^{-9}~\text{eV}}\right) \left(\frac{10^{-8}~\text{eV}}{m_a}\right)^{1/2} \left(\frac{F_a}{10^{14}~\text{GeV}}\right)^2\,,
\end{equation}
where $\theta$ denotes the initial misalignment angle of the axion field $a$, the parameter $\beta'$ quantifies the axion-dark photon coupling in $\frac{\beta'}{4F_a} a F^\prime_{\mu \nu} \tilde{F}^{\prime \mu \nu}$, $m_X$ and $m_a$ represent respectively the masses of the dark photon and axion, and $F_a$ is the axion decay constant.
Any production mechanism involving tachyonic instabilities predominantly produces a specific dark photon helicity, depending on model parameters. 

Other production mechanisms also include dark Higgs dynamics~\cite{Sato:2022jya,Redi:2022zkt}, and cosmic strings associated with the spontaneous breaking of the dark gauge symmetry~\cite{Long:2019lwl,Kitajima:2022lre}.

While dark photons can be produced through various mechanisms with a particular polarization within the cosmological horizon, the influence of DM halo formation as well as structure formation on the distribution of dark photon polarizations at megaparsec scales remains uncertain.
It is conceivable that the process of structure formation, where inhomogeneities in the gravitational potential can affect the dark photon equation of motion in the expanding Universe, may change the dark photon polarization~\cite{Arias:2012az}. 

Aside from this, it is expected that the polarization of the dark photon field can remain unchanged over cosmological time scales.
An estimate in Ref.~\cite{Caputo:2021eaa}, taking into account gravitational effects on the dark photon polarization, shows that the dark photon DM can preserve some degree of its initial polarization over the lifetime of a galaxy. 
While more dedicated studies will be needed to resolve this issue, the dark photon DM with a fixed polarization over typical experimental time scales seems quite plausible. 

In this work, we consider two limiting scenarios for dark photons comprising the entire DM abundance: random polarization and fixed polarization. 
Any realistic scenario will presumably lie between these two extremes.

\section{Appendix B: Data analysis procedure}
The TASEH experimental data are processed in the following way:
\begin{itemize}
\item The signal from the cavity is amplified and directed to the vector signal transceiver (VST) input for data acquisition, IQ demodulation and sampling, and the sampled IQ quadrature time trace data are sent to a network attached storage (NAS). 
A fast Fourier transform (FFT) is performed offline in real time to retrieve the frequency-domain power spectrum. 
The evaluated FFT signal power spectrum has a 2-MHz frequency bandwidth with a spectral resolution bandwidth $\Delta f_c  = 1$~kHz, sufficiently small to resolve an expected axion signal of 5~kHz width. 
To avoid aliasing effects, the VST rolls off the Intermediate Frequency (IF) signal via a band-pass filter, imposing a useful frequency span of 1.6~MHz. 
\item A Savitzky-Golay (SG) filter is applied to the averaged spectrum of each frequency scan by fitting adjacent points of successive sub-sets of data with a 4th-order polynomial, using a window width of 201 frequency bins. 
In the absence of a signal, the SG filter output can be considered as the averaged noise power. 
The raw averaged power spectrum is divided by the output of the SG filter, then a unity is subtracted from the ratio to get the dimensionless normalized spectrum, called the Relative Deviation of Power (RDP), denoted by $\delta^{\text{norm}} = \text{(Raw Power)}/\text{(SG filter)} -1$. 
The normalized spectrum from each scan is further rescaled such that a KSVZ axion signal will have a rescaled RDP $\delta^{\text{res}}_{ij}$ that is approximately equal to unity, if the signal power is contained within only one frequency bin:
\begin{align}
\delta^{\text{res}}_{ij}=R_{ij} \delta_{ij}, ~ \sigma^{\text{res}}_{ij}=R_{ij} \sigma_i ~,~  \text{for $i$-th scan, $j$-th bin}\,,
\end{align}
where 
\begin{align}
R_{ij}= \frac{k_B T_{\text{sys}} \Delta f_{\text{bin}}}{P^{\text{KSVZ}}_{ij} L(f_{ij},f_{ci},Q_L)}\,,
\end{align}
with $f_{ci}$ being the cavity frequency of $i$-th scan. 
The system-noise temperature $T_{\text{sys}} \sim T_a$ to a good approximation, where
the amplifier noise temperature $T_a$, which is a function of frequency, can be estimated from data fitting. 

\item During data taking, each frequency is scanned $15 \sim 16$ times. 
To combine spectra from different resonant-frequency scans, in order to obtain a single overall spectrum, one needs to apply a weighting algorithm. 
So, for a given frequency $f_{n}$, we define the combined spectrum as 
\begin{align}
\delta^{\text{com}}_n =\frac{\sum ( \delta^{\text{res}}_{ij} \cdot \omega^{ij}_n)}{ \sum  \omega^{ij}_n}, \quad\sigma^{\text{com}}_n = \frac{\sqrt{\sum (\sigma^{\text{res}}_{ij} \cdot \omega^{ij}_n )^2} }{\sum \omega^{ij}_n} \,, \label{eq:combrdp}
\end{align}
where the weight $\omega^{ij}_n = \frac{1}{ (\sigma^{\text{res}}_{ij} )^2}$, and the summations are over the $i$-th scan and $j$-th bin that matches the target frequency $f_{n}$. 
In practice, the summation over $i$ runs from 1 to 839~\footnote{There are 75 extra scans during the rescan stage.} (steps), the summation over $j$ runs from 1 to 1600 (bins), and the summation over $n$ runs from 1 to 92243 (bins). 
Note that the uncertainty of the averaged power in the overlapped regions of frequency space is reduced due to this combination.

\begin{figure}[htbp]
\begin{center}
\includegraphics[width=0.45\textwidth]{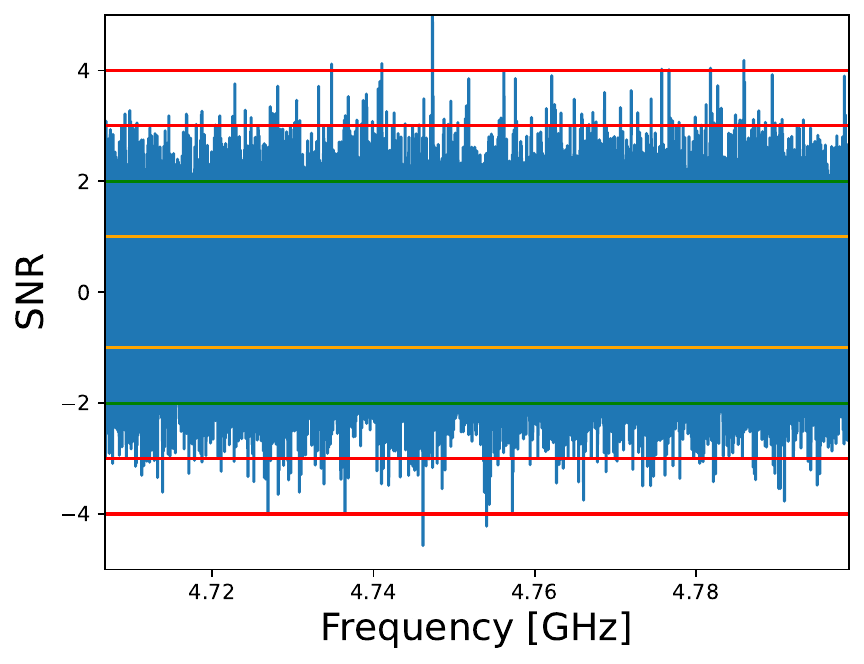}
\includegraphics[width=0.45\textwidth]{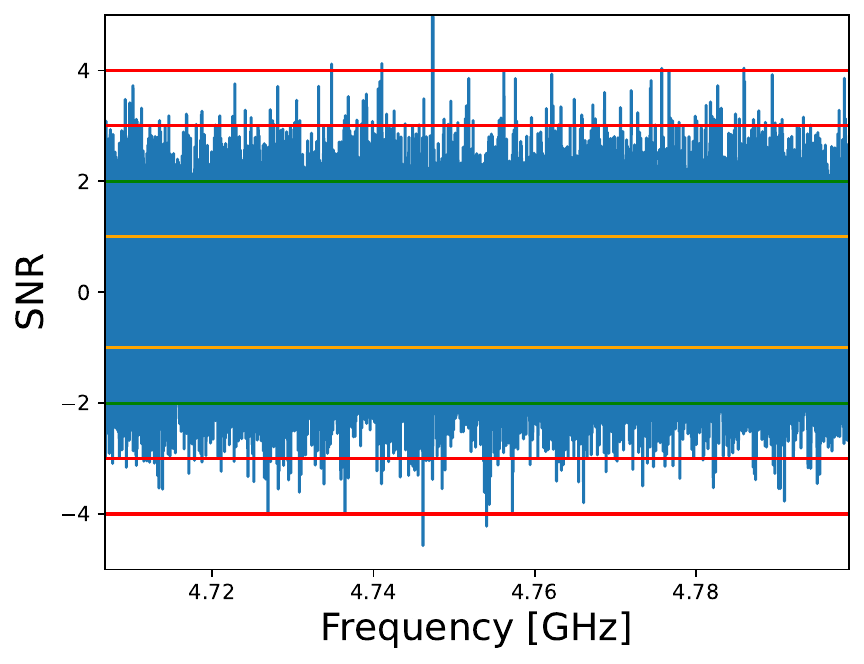}
\end{center}
\caption{The SNR for the combined spectrum before (left panel) and after (right panel) rescanning. 
\label{fig::combsnr}}
\end{figure}

Figure~\ref{fig::combsnr} shows the SNR of the combined spectrum before and after rescans. 
In total, 75 rescans were performed on frequencies whose merged SNR ({defined by $\delta^{\rm merge} / \sigma^{\rm merge}$ using the following Eq.~\eqref{eq:merge}}) exceeded 3.355, to assess if it was a real signal or a statistical fluctuation. 
As a result, we can observe that the two spectra in Figure~\ref{fig::combsnr} are similar to each other, except that the post-rescan data have fewer frequencies with large SNR excesses. 

\item The expected axion bandwidth is about 5~kHz at the frequency of $\sim 5$~GHz. 
In order to maximize the SNR, a running window of five consecutive bins in the combined spectrum is applied, and the five bins within each window are merged to construct a final spectrum. 
According to the signal lineshape $\mathcal{F}(f,f_X)$ ({as defined in the following Eq.~\eqref{eq:signalshape}}), the contributing bins need to be rescaled to have the same RDP, of which the standard deviation is used to define the maximum likelihood (ML) weight for merging. 
The rescaling factor can be calculated as 
\begin{align}
L_k = \int^{f_X +k \Delta f_c}_{f_X+(k-1) \Delta f_c} \mathcal{F}(f,f_X) \,df, ~\text{ for } k=1,\cdots, 5\,,
\end{align}
where $f_X$ is the dark photon frequency. 
In practice, instead of using an $L_k$ that depends on the frequency $f_X$, the average ($\bar{L}_k$) over the range of $f_X$ is used:
\begin{align}
\bar{L}_k = 0.23, 0.33, 0.21, 0.11, 0.06 ~\text{ for } k=1,\cdots, 5\,.
\end{align}
Note that the relative variation of $f_X$ is at most 90~MHz$/$5~GHz $\simeq$ 2\% and the lineshape of $\mathcal{F}(f,f_X)$ can be considered constant for the full range of the operational frequency. 
Therefore, the value of $L_k$ has only a weak dependence on $f_X$. 
The RDP and the standard deviation of the merged spectrum can then be calculated from the combined spectrum via  
\begin{align}
\delta^{\text{merge}}_g = \frac{\sum_k \frac{\delta^{\text{com}}_{g+k-1}}{\bar{L}_k} \cdot \left(\frac{\bar{L}_k}{\sigma^{\text{com}}_{g+k-1}}\right)^2}{\sum_k \left(\frac{\bar{L}_k}{\sigma^{\text{com}}_{g+k-1}}\right)^2}\,,\quad \sigma^{\text{merge}}_g = \frac{1}{\sqrt{\sum_k \left(\frac{\bar{L}_k}{\sigma^{\text{com}}_{g+k-1}}\right)^2}} \,. \label{eq:merge}
\end{align}

\begin{figure}[htbp]
\begin{center}
\includegraphics[width=0.45\textwidth]{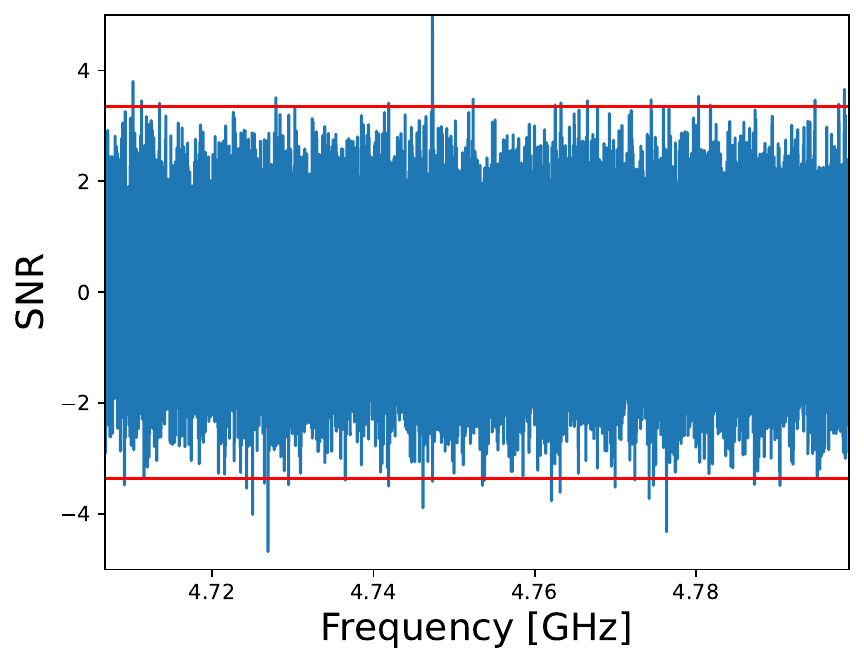}
\includegraphics[width=0.45\textwidth]{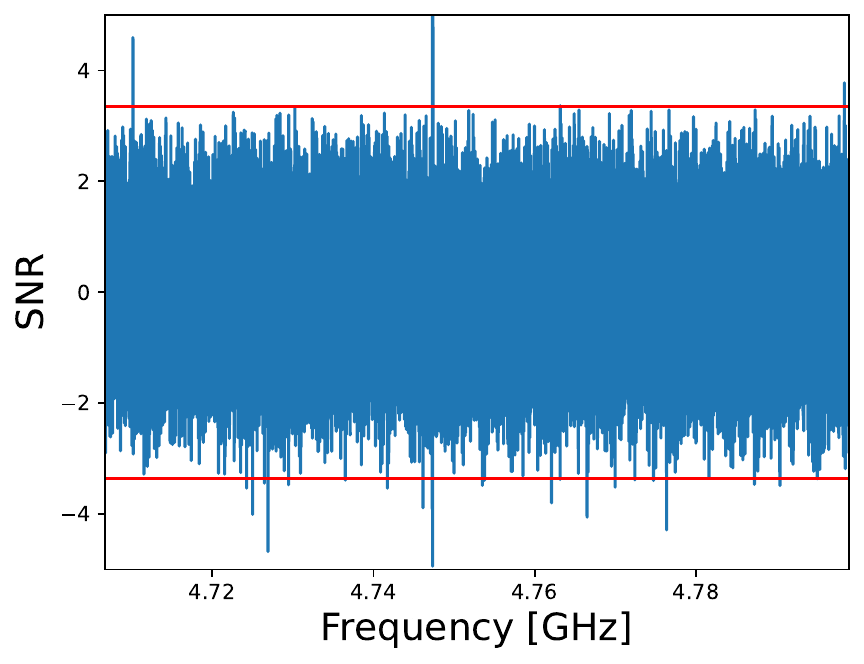}
\end{center}
\caption{The signal-to-noise ratio calculated for the merged spectrum before (left panel) and after (right panel) the rescan.
\label{fig::mergesnr}}
\end{figure}

Figure~\ref{fig::mergesnr} shows the SNR of the merged spectrum before (left panel) and after (right panel) the rescan. 
During the main scan, 22 candidates with an SNR greater than 3.355 were found. 
Among them, 20 candidates disappeared after rescanning, and two candidates remained.
The signal in the $4.74730 – 4.74738$~GHz frequency range was detected via a portable antenna outside the DR and judged to have originated from the instrument control computer in the laboratory. 
The signal in the $4.71017 – 4.71019$~GHz frequency range was not detected outside the DR, but was still present after turning off the external magnetic field, implying that it did not arise from the axion DM.

\item Aside from these two frequency ranges, no candidates were found after rescanning. 
An upper limit on the signal power $P_s$ is derived by setting $P$ equal to $5 \sigma_{g}^{\text{merged}} \times P_{g}^{\text{KSVZ}}$, where $P_{g}^{\text{KSVZ}}$ is the expected signal power for the KSVZ axion for a certain frequency bin $g$ in the merged spectrum.
\end{itemize}

\subsection{Appendix C: Conversion factors for polarized dark photon DM}
Based on the details of the scanning strategy, we can calculate the conversion factor as defined in Eq.~(\ref{eq:convfactor}) for the TASEH experiment. 

We illustrate the calculation procedure at an arbitrarily chosen frequency $f=4.712705$~GHz, which was scanned 15 times during data taking.
The period and some of the detector parameters for each scan are presented in Table~\ref{tab:time4712705}.  
We can see that the signal power was highest during the 7th and 8th scans, when the cavity resonance was closest to $4.712705$~GHz. 
Meanwhile, the signal power was highly suppressed by the Lorentzian cavity response (as indicated in the last column) during the first and last few scans. 

\begin{table}[htbp]
	\centering 
	\scalebox{1}{
	\begin{tabular}{|c|c|c|c|c|c|c|c|c|} \hline 
 \multirow{2}{*}{Number}&    \multicolumn{2}{|c|}{Start} & \multicolumn{2}{|c|}{End} & \multirow{2}{*}{Cav. Freq. [GHz]} &  \multirow{2}{*}{$Q_0$} &  \multirow{2}{*}{Lorentzian response} \\ \cline{2-5}
   & Date & Time & Data & Time & & & \\ \hline
 1   &  2021.11.13 &   19:24.49    &    2021.11.13  & 20:06.58 &  4.713403      &    64662 & 0.02395049 \\ \hline
 2   &  2021.11.13  & 20:10.45    &    2021.11.13 &  20:52.54  & 4.713293      &    64863  & 0.03322079\\ \hline 
 3  &   2021.11.13  & 21:00.27   &     2021.11.13 &  21:42.37  & 4.713188     &     64963 & 0.04831467 \\ \hline 
 4 &    2021.11.13  & 21:45.38     &   2021.11.13  & 22:27.48   &4.713079     &     64846 &  0.07831831\\ \hline 
 5 &    2021.11.13  & 22:30.56   &     2021.11.13  & 23:13.06   &4.712975     &     64778 & 0.14043321 \\ \hline 
 6  &   2021.11.13  & 23:17.56    &    2021.11.14  & 00:00.05  & 4.712874      &    65002 & 0.29284697 \\ \hline 
 7 &    2021.11.14  & 00:04.25    &    2021.11.14 &  00:46.36  & 4.712771     &     64858 & 0.73170354 \\ \hline 
 8 &    2021.11.14 &  00:49.28     &   2021.11.14  & 01:31.38  & 4.712664     &     64654 & 0.87671718 \\ \hline 
 9  &   2021.11.14  & 01:34.39   &     2021.11.14  & 02:16.48  & 4.712561      &    65078  & 0.3626465 \\ \hline 
10 &     2021.11.14  & 02:19.19   &     2021.11.14  & 03:01.29  & 4.712459     &     65090 & 0.16309954  \\ \hline 
 11 &    2021.11.14 &  03:03.43    &    2021.11.14  & 03:45.53 &  4.712352     &     64976 &0.08673592  \\ \hline 
 12 &    2021.11.14  & 03:48.02   &     2021.11.14  & 04:30.11  & 4.712249      &    64886 & 0.05398879  \\ \hline 
  13&    2021.11.14  & 04:32.47    &    2021.11.14 &  05:14.57 &  4.712140     &     65058  & 0.03565758 \\ \hline 
  14  &   2021.11.14  & 05:18.40   &     2021.11.14 &  06:00.51 &  4.712038    &      64994 & 0.02589452  \\ \hline 
  15&    2021.11.14 &  06:04.33    &    2021.11.14  & 06:46.43  & 4.711938     &     64922 & 0.01974897 \\ \hline 
    
 	\end{tabular}}
	\caption{\label{tab:time4712705} The scanning periods, cavity frequencies, unloaded quality factors, and the corresponding Lorentzian responses for 15 scans at $f=4.712705$~GHz. }
\end{table}

\begin{figure}[htbp]
\begin{center}
\includegraphics[width=0.48\textwidth]{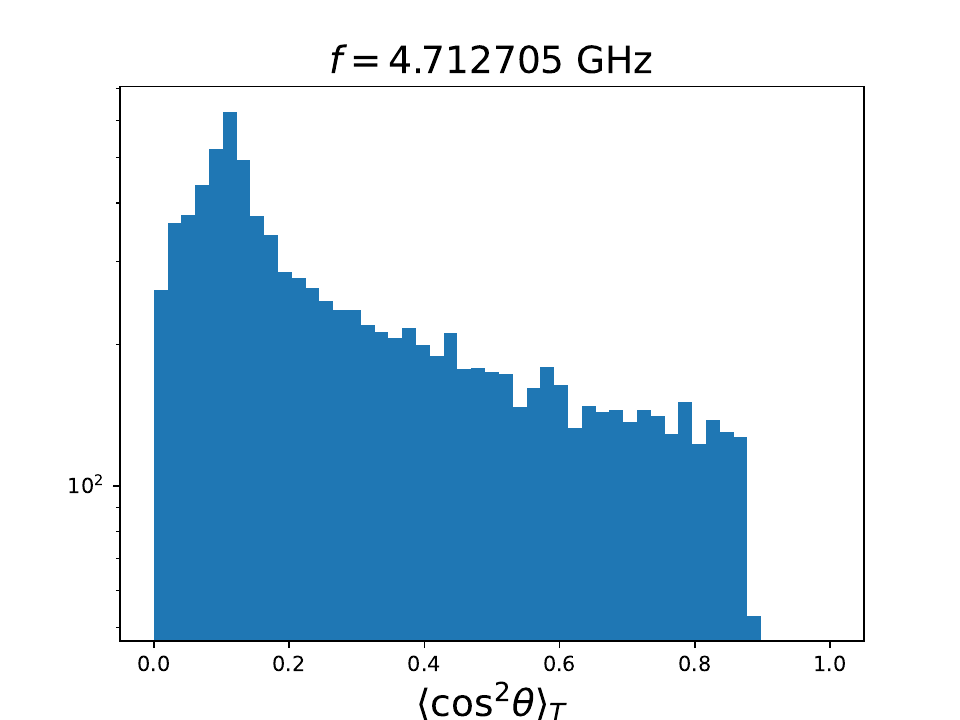}
\includegraphics[width=0.48\textwidth]{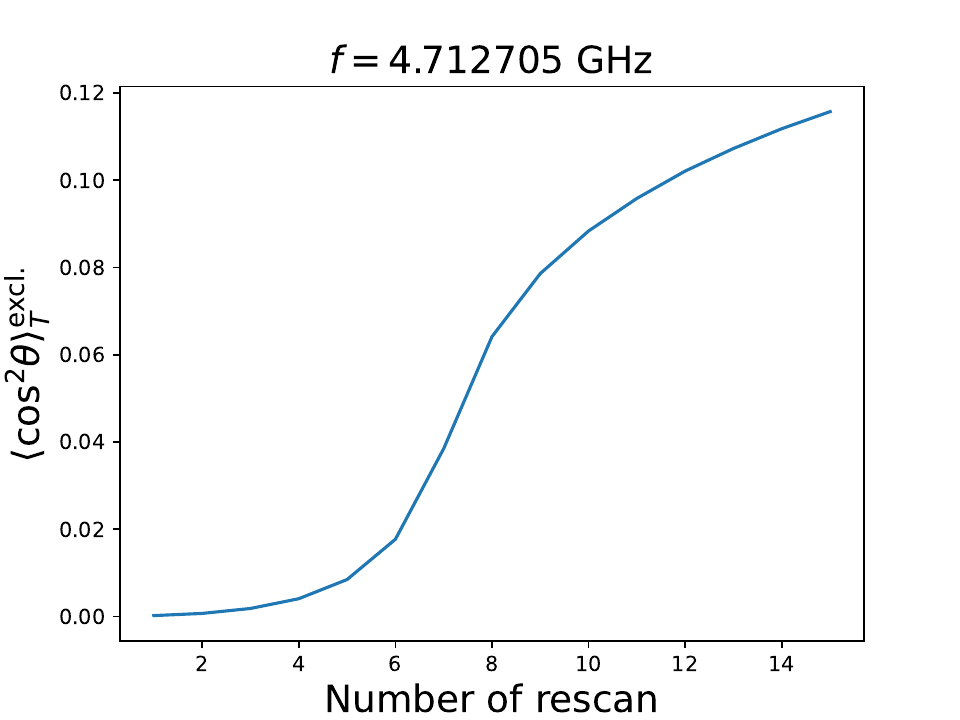}
 \end{center}
\caption{Left panel: The distribution of $\langle \cos^2 \theta(t) \rangle_T$ for the frequency $f=4.712705$~GHz with the TASEH scanning strategy. 
Right panel: The evolution of the conversion factor for $f=4.712705$~GHz with increasing numbers of scans. 
\label{fig::cost}}
\end{figure}

Using the measurement timing information as well as the signal power, we can infer the distribution of $\langle \cos^2 \theta(t) \rangle_T$ via isotropic sampling of $(\theta_X, \phi_X)$ in Eq.~\eqref{eq:cost2}. 
For the frequency $f=4.712705$~GHz, this distribution is given in the left panel of Fig.~\ref{fig::cost}. 

The distribution of $\langle \cos^2 \theta(t) \rangle_T$ is used in Eq.~\eqref{eq:conv1} to calculate the 95\% C.L. limit for the conversion factor. 
As multiple measurements not only increase the statistics but also permit the sensitivity to a wider range of dark photon polarization directions, one would expect that the conversion factor will increase with more scans. 
This can be seen in the right panel of Figure~\ref{fig::cost}, where we plot the evolution of the conversion factor for $f=4.712705$~GHz with increasing numbers of scans. 
As the signal power is highly suppressed by the Lorentzian response factor for the first and last few scans, their contribution to the conversion factor is mild. 
The dominant contribution comes from the 6th-10th scans. 

\begin{figure}[htbp]
\begin{center}
\includegraphics[width=0.5\textwidth]{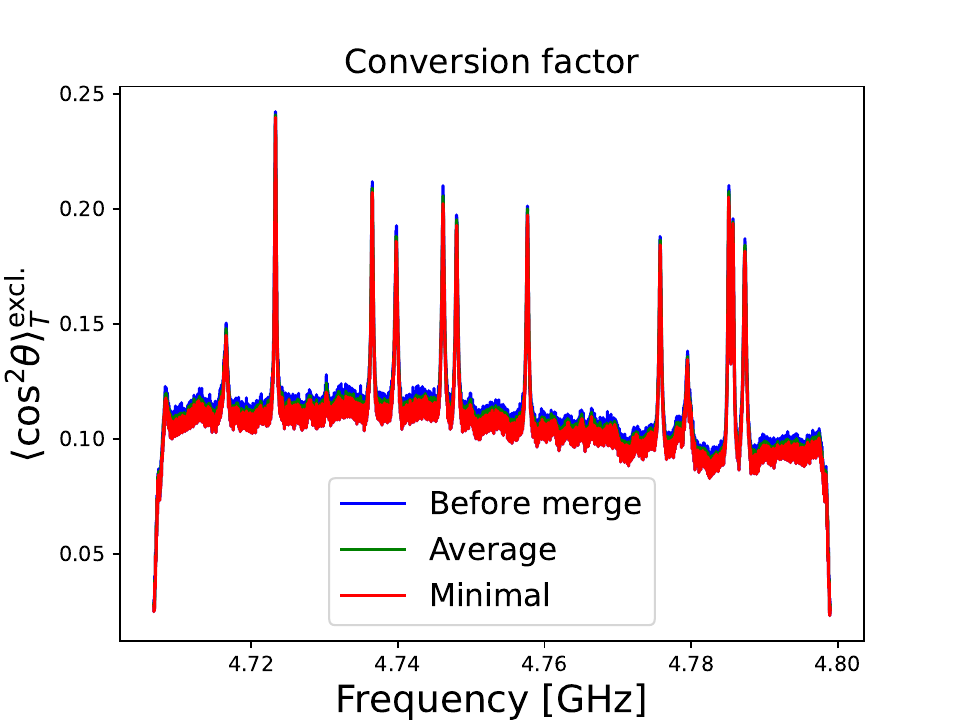}
 \end{center}
\caption{The conversion factor for the full frequency range at the TASEH experiment. 
The blue, green, and red curves correspond respectively to the conversion factor prior to bin merging, and the conversion factors post bin merging by taking either the average or minimum conversion factor among five adjacent bins. 
\label{fig::convfactor}}
\end{figure}

The above procedure is applied to all frequencies to calculate the full conversion factor. 
As presented in Figure~\ref{fig::convfactor}, the typical value of the conversion factor is mainly attributed to the latitude of TASEH detector at 25$^{\circ}$ as well as the total measurement time of a few hours. 
There were some cases in which the time intervals among scans were much longer than a few minutes, creating peaks in the conversion factor, such as at $f\sim$ 4.787~GHz. 
There were also other scans with integration times much longer than 40 minutes, which also had improved conversion factors. 

In principle, the merging procedure for merging adjacent bins that may have been measured at different times can have some effects on the conversion factor. 
However, since the measurement times of frequencies in a single scan are the same, merging these frequencies will not lead to any modifications. 
Therefore, we only need to consider the change in conversion factors for the frequencies at the edges of each scan.

In this case, we can compute the conversion factor for the post-merge data by calculating the weighted sum of the conversion factors of five consecutive bins. 
To illustrate the effects of this, we present in Figure~\ref{fig::convfactor} the conversion factors before and after merging 
by taking either the average or the minimal conversion factor among the five consecutive bins, respectively. 
We can observe that the effects of the merging procedure are small over the full frequency range. 

\section{Appendix D: Fitting procedure on the signal excess}
\begin{figure}[htbp]
\begin{center}
\includegraphics[width=0.45\textwidth]{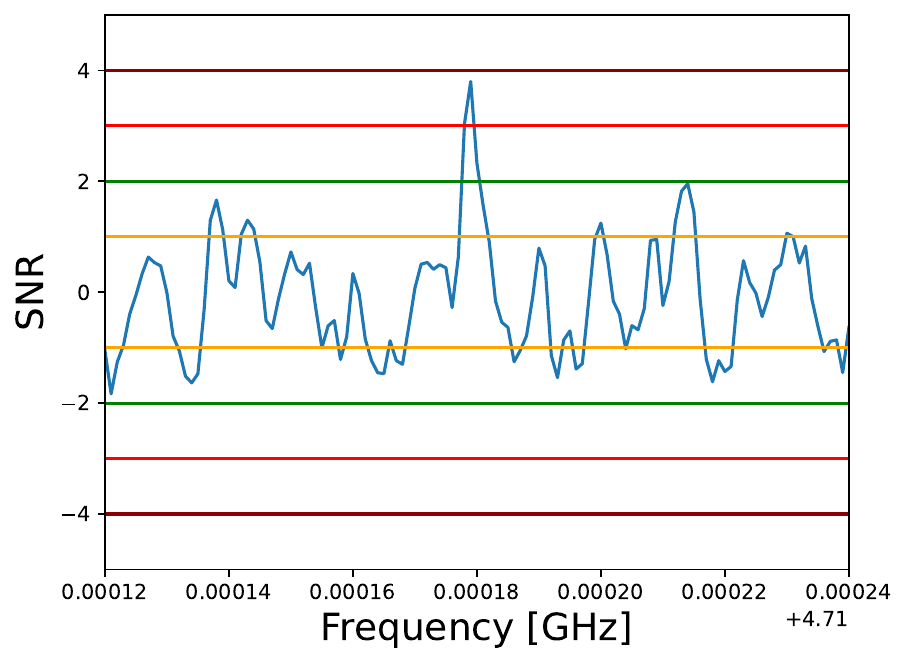}
\includegraphics[width=0.45\textwidth]{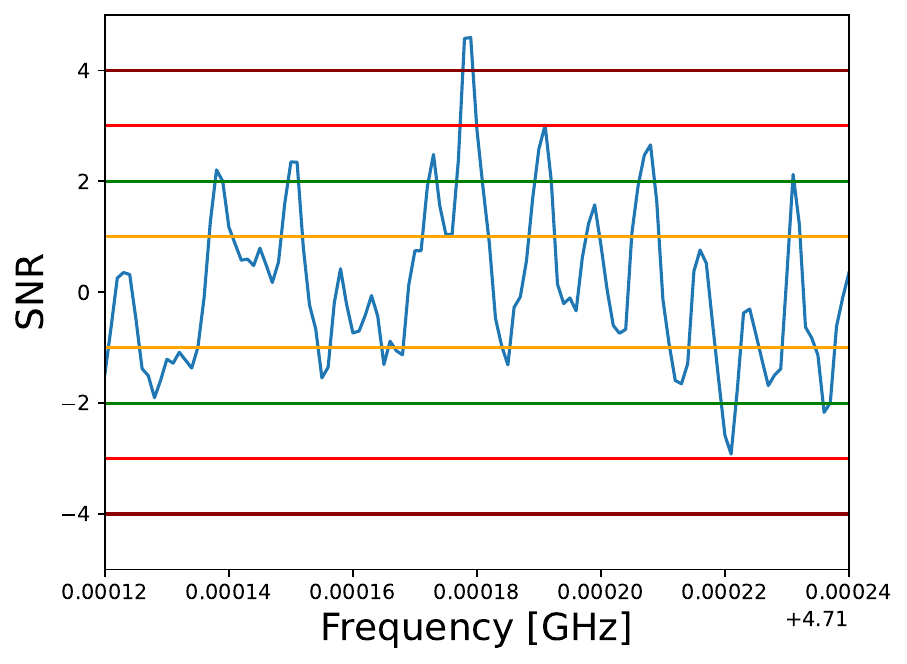}
\end{center}
\caption{
The signal-to-noise ratio for the merged spectrum before (left panel) and after (right panel) the addition of rescan data in the frequency range of $[4.710120,4.710240]$~GHz. 
\label{fig::snr710}}
\end{figure}

In Figure~\ref{fig::snr710}, we present the SNR for the merged spectrum before and after the addition of rescan data in the frequency range of $[4.710120,4.710240]$~GHz. 
The SNR is highest at $f=4.710179$~GHz, and it increases from 3.801 to 4.593  after rescans. 
The frequency width of the signal is a few kHz, consistent with what would be expected from a dark photon signal.

The lineshape of the signal can be more easily seen in the RDP before applying the merging procedure.  Here we adopt the rescaled RDP as defined in Eq.~\eqref{eq:combrdp} for the signal shape analysis. 
Note that one unit in $\delta^{\text{res}}$ for each frequency bin refers to a KSVZ axion signal with power contained entirely in that bin. 
As a result, the dark photon signal power (assuming random polarization $\langle \cos^2 \theta(t) \rangle_T =1/3$) is normalized to the KSVZ axion signal power with a scaling factor
\begin{align}
P_{X} =\left( \frac{\epsilon}{\epsilon_0} \right)^2 P_{\text{KSVZ}} = \left( \frac{\epsilon}{\epsilon_0} \right)^2 \times \left(g^2_{a\gamma\gamma} \frac{\rho_a}{m_a^2} \omega_c B^2_0 V C_{010} Q_L \frac{\beta}{1+\beta} \right)\,,
\end{align}
where the normalized kinetic mixing is
\begin{align}
\epsilon_0 = \frac{g^{\text{KSVZ}}_{a\gamma \gamma} B_0}{m_a \cos \theta} \sim 1.005 \times 10^{-15}\,.
\end{align}

Moreover, the spectral shape of a haloscope signal is proportional to the dark photon kinetic energy distribution. 
Assuming that the dark photon DM velocity distribution follows a Maxwell-Boltzmann distribution in the galactic rest frame, with the local density $\rho = 0.45$ GeV/cm$^{3}$ and the local circular velocity $v_c=220$ km/s, the kinetic energy distribution is given by
\begin{align}
   F(f) = 2&\left(\frac{f - f_{X}}{\pi}\right)^{1/2}\left(\frac{3}{ f_{X} \beta_{DM}^2}\right)^{3/2}  \exp\left(-\frac{3(f - f_{X})}{f_{X} \beta_{DM}^2}\right) \,, \label{eq:signalshape}
\end{align}
with $\beta^2_{DM} =\langle v^2 \rangle / c^2=(270~\text{km/s})^2/(3\times 10^5 ~\text{km/s})^2$. 
In the Lab frame, accounting for the orbital velocity of the Sun, the spectrum of dark photon DM can be well approximated by replacing $\beta^2_{DM}$ with $1.7 \beta^2_{DM}$ in the equation above~\cite{Brubaker:2017rna}. 



\end{document}